\documentclass[aps,twocolumn,graphics,floatfix,tightenlines,amsmath,amsymb,]{revtex4-2}
\usepackage{graphics}
\usepackage{epstopdf}
\usepackage{epsfig}
\usepackage{graphicx}
\usepackage{epsf,epic}
\usepackage{color}
\usepackage{subfig}
\usepackage{amsmath}
\usepackage{booktabs}
\usepackage{multirow}
\usepackage{physics}
\usepackage{hyperref}
\usepackage{amsfonts}
\usepackage{wrapfig}
\usepackage{breqn}
\usepackage{pstricks}
\usepackage[utf8x]{inputenc}
\usepackage{multirow}
\usepackage{fancyref}
\usepackage{pst-node}
\usepackage{float}
\usepackage{bm}
\usepackage{dcolumn}
\newcommand{\etal}{\textit{et al.\ }}

\makeatletter
\usepackage{etoolbox} 
\appto{\appendix}{%
	\@ifstar{\def\theequation@prefix{A.}}%
	{}%
}
\preto\maketitle{%
  \begingroup\lccode`~=`,
  \lowercase{\endgroup
  \let\saved@breqn@active@comma~
  \let~}\active@comma 
}
\appto\maketitle{%
  \begingroup\lccode`~=`,
  \lowercase{\endgroup
  \let~}\saved@breqn@active@comma 
}
\makeatother

\begin{document}
\title{Symmetry breaking structural relaxation and optical transitions of native defects and carbon impurities in LiGa$_5$O$_8$}
\author{Klichchupong Dabsamut}\email{klichchupong.dab@cra.ac.th}
\affiliation{Department of Science, Technology and Innovation, Faculty of Science, Chulabhorn Royal Academy, Bangkok 10210, Thailand}
\author{Adisak Boonchun}
\affiliation{Department of Physics, Faculty of Science, Kasetsart University, Bangkok 10900, Thailand}
\author{Walter R. L. Lambrecht}\email{walter.lambrecht@case.edu}
\affiliation{Department of Physics, Case Western Reserve University, 10900 Euclid Avenue, Cleveland, Ohio 44106-7079, USA}
\begin{abstract}
LiGa$_5$O$_8$ in a spinel type structure has recently been claimed to be an unintentional p-type ultra-wide-band-gap oxide semiconductor. While previous computational work did not yet identify the origin of p-type doping and in fact predicted insulating behavior by compensation of deep acceptors by shallow donors,  defect characterization  in terms of its optical signatures remains important.  Rather than focusing  on thermodynamics transition levels, as in earlier work, this present paper focuses on 
  the vertical transitions in a defect configuration diagram of defects in different charge states,  representing absorption  and emission processes involving carrier capture/emission  from/to band edges. 
In addition, the structural relaxation of several native defects is revisited by allowing for more complex symmetry-breaking distortions in an effort to reconcile conflicting results in the previous literature. Special attention is given to the Li vacancy because it is the shallowest native acceptor. For this defect, the previously reported transition levels are revised on the basis of symmetry-breaking relaxations. The structural relaxations, band structures, and densities of states are compared between the symmetry-broken polaronic and symmetry-conserving non-polaronic states. Finally, we also study carbon impurities, which are likely to originate from growth methods involving organic precursors. \end{abstract}
\maketitle
\section{Introduction}
Cubic spinel-type structured LiGa$_5$O$_8$ has recently attracted attention as an  ultra-wide-band-gap (UWBG) semiconductor, which unexpectedly features unintentional p-type doping \cite{Zhao2023,Kaitian25,Vangipuram25}. Its band gap obtained from a Tauc-plot was reported to be 5.36 eV, in good agreement with quasiparticle self-consistent (QS)$GW$ calculations \cite{Lambrecht2024} which find a somewhat larger direct quasiparticle fundamental gap of 5.84 eV and  slightly lower indirect  gap of 5.74 eV but also a strong exciton binding energy, resulting in an estimated optical gap of 5.2$\pm$0.2 eV. Our previous calculations of native defects and various complexes \cite{Dabsamut24} were thus far unable to explain the p-type doping as all candidate acceptors were found to have deep acceptor binding energies and furthermore, in equilibrium the Fermi level would be pinned deep in the gap by the compensation of acceptors by compensating donors, mainly Ga$_{\rm Li}$ which is a shallow donor and Oxygen vacancies which are deep donors. Experimentally\cite{Kaitian25}, it was indeed found that annealing in oxygen rich atmosphere leads to insulating behavior and p-type behavior occurs under O-poor and  Ga-rich conditions.
The shallowest acceptor found in our previous  work is the $V_{\rm Li}$ with a $0/-1$ transition level at 0.74 eV above the Valence Band Maximum (VBM). A paper by Lyons \cite{Lyons24} reached similar conclusions and emphasized the polaronic nature of the acceptors. However, it reported an even deeper energy level for the $V_{\rm Li}$ at 1.25 eV with the hole localized on a single O neighbor, while our previous calculations indicated the hole to be spread over the six neighboring oxygens of the octahedral environment of Li. In this paper we therefore revisit the Li-vacancy and we will indeed show that symmetry breaking relaxation results in a deeper polaronic state. Our results also deviate from Lyons' for the Ga-vacancies,
where we found a single $0/3-$ transition level, whereas he found separate $0/1-$, $1-/2-$, $2-/3-$ states. We will show that this also can be resolved by including lower symmetry configurations and finding the global minimum more accurately. 

At the same time this suggests that the $V_{\rm Li}$ could have a dual nature with a shallower non-polaronic  and a deeper polaronic configuration. This type of scenario was recently suggested as a possible explanation for shallow donor and acceptor behavior in AlN \cite{YujieLiu25}, another UWBG  material, which was found extremely challenging to dope. While our non-polaronic $V_{\rm Li}$ is still too deep to explain p-type doping, further study of this defect seems warranted. In particular besides the standard approach of calculating the transition levels, it is worthwhile examining its band structure in more detail, which we are doing in this paper. While this explains the difference between the polaronic and non-polaronic state, the non-polaronic state has higher total energy and is not a local minimum separated by a barrier from the polaronic state and furthermore is still too deep to provide a source of p-type doping.   

In our previous work, we considered also donor acceptor pairs but these all had net donor character. Here, we extend this search for a shallow acceptor by considering a $V_{\rm Li}$ next to a Ga-Li exchange defect. In other words, we combine the double donor Ga$_{\rm Li}$ with a double acceptor Li$_{\rm Ga}$ and a single acceptor $V_{\rm li}$. While this can in principle then lead to a net acceptor state, we find instead that it is an amphoteric defect with the  donor character dominating near the VBM. We may also wonder about impurities as possible sources
of the p-type doping. Since carbon is a common impurity introduced in metalorganic chemical vapor deposition (MOCVD)
or in mist chemical vapor deposition (CVD) from the precursors,\cite{meng2023role,hernandez2021mocvd,vasin2025challenges} we study the site selectivity and defect levels of carbon impurities.
However, none of them is found to be a shallow acceptor.

Even if an as yet unknown  shallow acceptor could be found or introduced in large enough concentration by doping, we note that the energy of formation of the native
shallow donor Ga$_{\rm Li}$ has large negative energy of formation if the Fermi level is near the valence band maximum and will thus necessarily lead to compensation and thereby push the equilibrium Fermi level deeper into the gap. 
The above results indicate that intrinsic LiGa$_5$O$_8$ cannot be p-type in equilibrium and that observations of p-type behavior must result  from some non-equilibrium situation in part of the sample, perhaps near extended defects.  Nonetheless it is important to characterize the defect physics experimentally and to do this deeper insight is required in the
optical signatures of the defects. 
The main objective of this paper is therefore to assist in the identification of defects by optical experimental probes such as photoluminescence or cathodoluminescence.

The remainder of the paper is organized as follows. After briefly describing the methodology used
in Sec.\ref{sec:method} we present the results for revised native defect transition levels in Sec. \ref{sec:native}, the results for a triple defect complex in
Sec.\ref{sec:complex} and for C impurities in Sec. \ref{sec:carbon}. We then focus on the  optical transitions between
band edges and the defect levels in Sec.\ref{sec:config} and discuss their relation to experiment in Sec.\ref{sec:expt}.
Finally, we summarize our main conclusions in Sec. \ref{sec:conclusion}.

 \section{Methods}\label{sec:method}
The calculations here were performed using the Vienna Ab-initio Simulation Package (VASP) \cite{VASP1,VASP2} code using the Projector augmented Wave (PAW)\cite{PAW} method and using the Heyd-Scuseria-Ernzerhof (HSE) hybrid functional\cite{HSE03,HSe06} with parameters chosen so as to reproduce the QS$GW$ indirect gap of $\sim5.72$  eV as documented in \cite{Dabsamut24}.
Other computational details are the same as in \cite{Dabsamut24}.To construct the defect configuration diagram for two charge states, we calculate the total energy of each charge state both in its own relaxed equilibrium geometry and in the relaxed geometry of the other charge state. Here we focus on the native defects  without considering the defect pair complexes. In our previously published results, 
all of these, except the $V_{\rm Ga-oct}$ have a single transition level in the gap. Depending on how that level is occupied we may consider transitions to both the VBM and CBM. 
\section{Results}\label{sec:results}
\subsection{Revision of transition levels}\label{sec:native}

\begin{figure}[h]
	\centering
	\includegraphics[width=6.5cm]{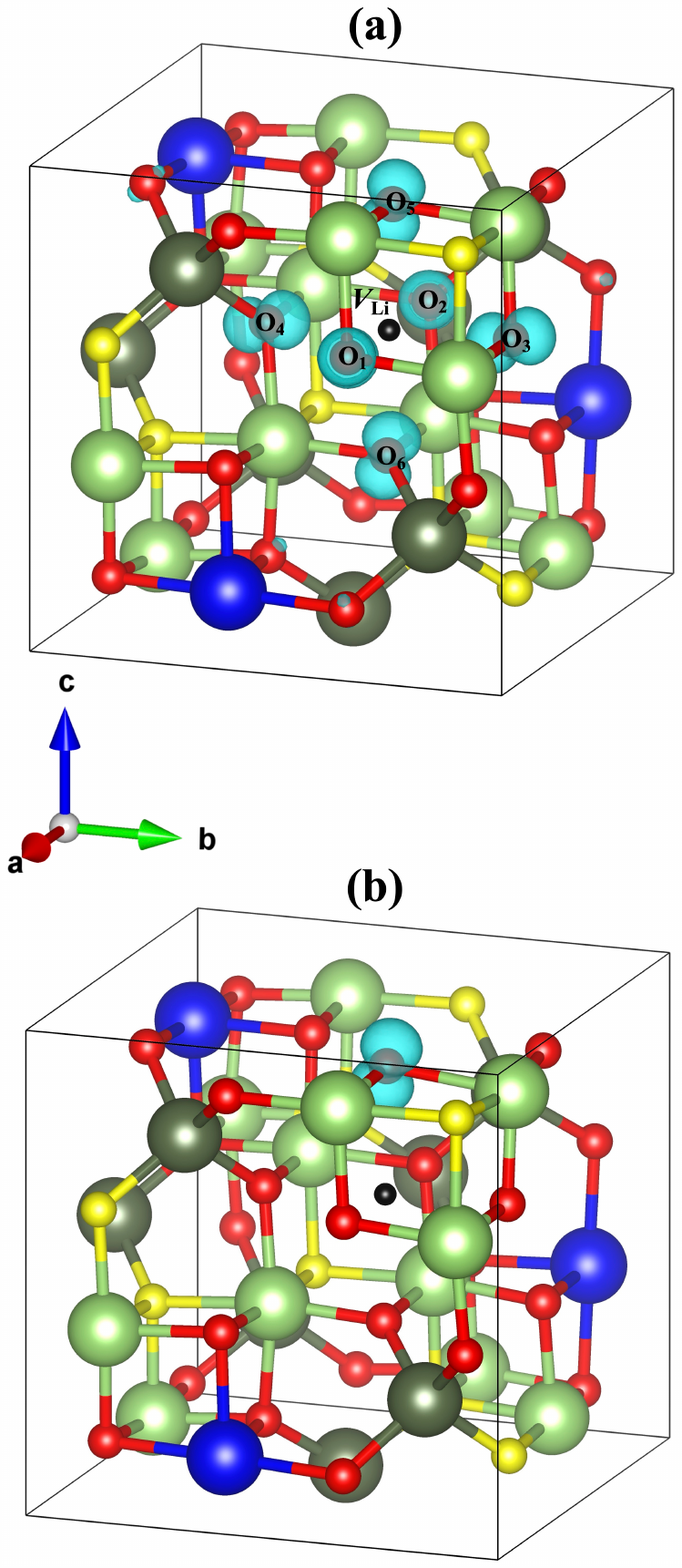}
	\caption{Local structure of the Li vacancy in LiGa$_5$O$_8$. (a) $V_{\rm Li}$ (symmetry), where the six neighboring O atoms remain nearly equivalent. (b) $V_{\rm Li}$ (symmetry-broken), showing a symmetry-lowering distortion of the O$_5$ environment and a localized defect state. The O atoms are labeled O1--O6 as defined in Table~\ref{tab:VLi_bondlengths}.}
	\label{fig:VLi_symbreak}
\end{figure}

As mentioned in the introduction we revised several calculations to allow for symmetry breaking distortions. 
We first discuss the $V_{\rm Li}$. We found that our previous calculations were restricted to symmetry preserving
relaxations. Displacing the atoms from their position in a symmetry breaking manner as a new starting point and carrying out further relaxations, we found a new symmetry-broken configuration with lower energy for the neutral charge state, which then leads to a deeper $0/-$ transition at 1.58 eV. Furthermore, in this configuration, the hole is localized on a single  oxygen as shown in Fig. \ref{fig:VLi_symbreak}. It is found to be even deeper than predicted by Lyons\cite{Lyons24} but consistent with  his finding of a polaronic single O nature of the defect wavefunction.

Table~\ref{tab:VLi_bondlengths} compares the O--$V_{\rm Li}$ distances in the ideal LiGa$_5$O$_8$ structure with those obtained for $V_{\rm Li}$ (symmetric) and $V_{\rm Li}$ (symmetry-broken). In the $V_{\rm Li}$ (symmetric) case all six neighboring oxygens relax to essentially the same distance (2.144~\AA), consistent with the hole being spread over the six O neighbors as found in our earlier work and corresponding to a 3\% outward relaxation.
We may further note that the oxygen atoms relax outward from the defect indicating that their bonding with the second shell of neighbors is increased to compensate for the loss of bonding to the central Li. In contrast, the $V_{\rm Li}$ (symmetry-broken) structure shows a clear splitting of the O--$V_{\rm Li}$ distances, indicating a symmetry breaking relaxation in the octahedral environment. Compared to the symmetrically relaxed state two $V_{\rm Li}$-O distances slightly decrease while three increase and the hole localizes on O6 which is about 5 \% outward relaxed from the ideal position.
The particular O on which the hole localizes could of course be any of the six neighbors depending on the type of deformation that occurs. The symmetry broken neutral charge state has 0.84 eV lower energy than the previous calculation and as the single negative  charge state stays the same this means that the transition level moves further above the VBM by 0.84 eV and now lies at 1.58 eV instead of previously 0.74 eV.

\begin{table}[h]
	\caption{O--$V_{\rm Li}$ distances (in \AA) for the ideal LiGa$_5$O$_8$ structure, $V_{\rm Li}$ (symmetry), and $V_{\rm Li}$ (symmetry-broken).\label{tab:VLi_bondlengths}}
	\begin{ruledtabular}
		\begin{tabular}{cccc}
			O--$V_{\rm Li}$ distance &  LiGa$_5$O$_8$ & symmetry &  symmetry-broken \\
			\hline
			O1 & 2.073 & 2.144 & 2.144 \\
			O2 & 2.073 & 2.144 & 2.129 \\
			O3 & 2.073 & 2.144 & 2.150 \\
			O4 & 2.073 & 2.144 & 2.126 \\
			O5 & 2.073 & 2.144 & 2.169 \\
			O6 & 2.073 & 2.144 & 2.163 \\
		\end{tabular}
	\end{ruledtabular}
\end{table}

\begin{figure}
	\centering
\includegraphics[width=6.5cm]{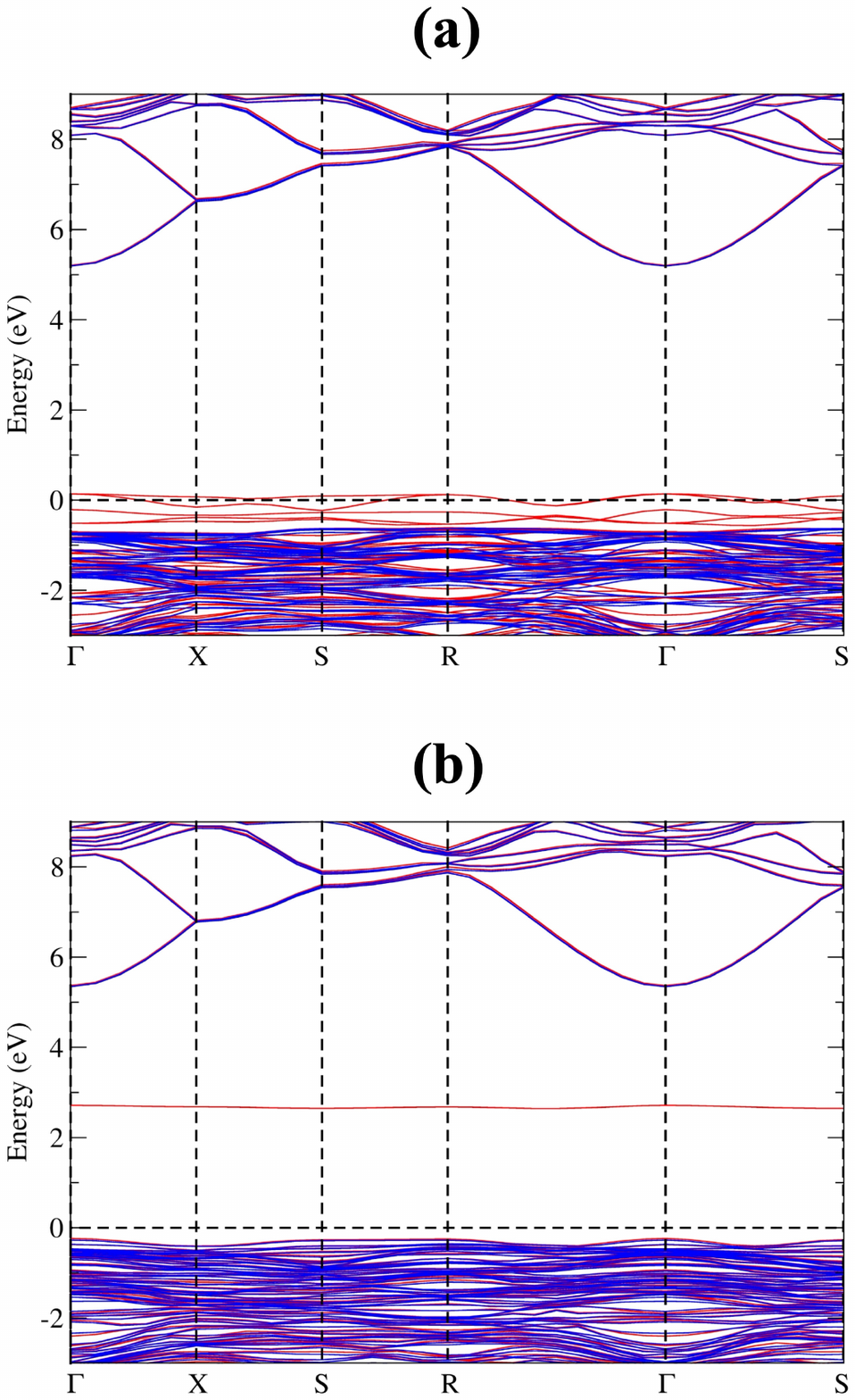}
  \caption{Band structure of LiGa$_5$O$_8$ with (a) symmetrically relaxed $V_{\rm Li}$ and (b) symmetry broken structure}\label{figbndvlisym} 
  \end{figure}
\begin{figure}
  \centering
  \includegraphics[width=6.5cm]{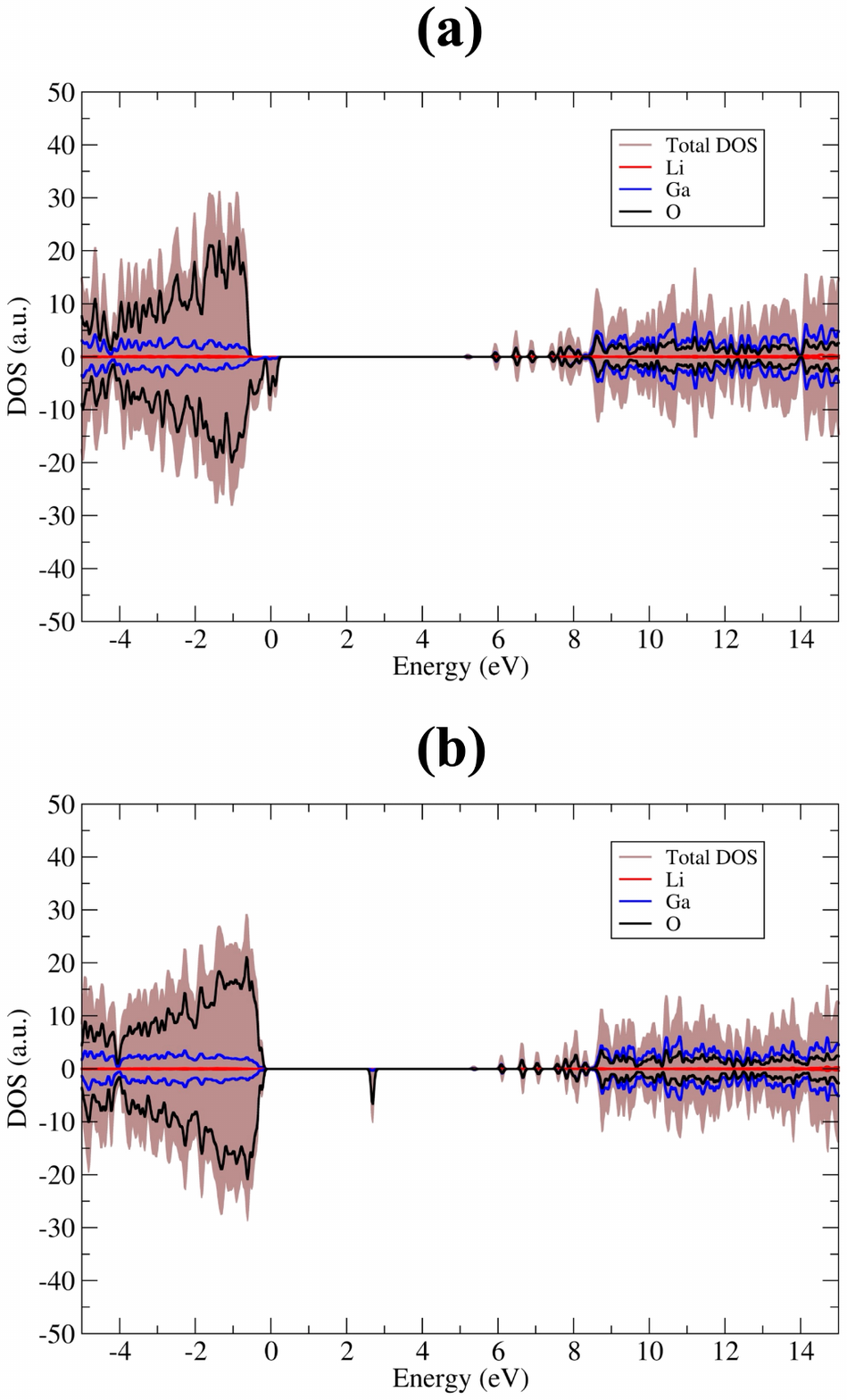}
  \caption{Partial and total densities of states for $V_{\rm Li}^0$ for (a), symmetric and (b) symmetry broken state}\label{figpdosvlisym}
\end{figure}
\begin{table}
  \caption{O-distances (in \AA) surrounding  $V_{\rm Ga-t}$ after full relaxation and from \cite{Dabsamut24} in parentheses.}\label{Tab-bonds-VGat}
  \begin{ruledtabular}
    \begin {tabular}{ccc}
      neighbor & $q=-1$ & $q=-2$ \\ \hline
      O1      & 2.039 (2.003) & 2.039 (2.011)  \\
      O2      &  2.058 (2.076) & 2.070 (2.077) \\
      O3      & 1.967 (2.003)  & 2.004 (2.011) \\
      O4      & 1.983 (2.003)  & 2.005 (2.011) \\
    \end {tabular}
  \end{ruledtabular}
\end{table}

\begin{figure}
  \includegraphics[width=6cm]{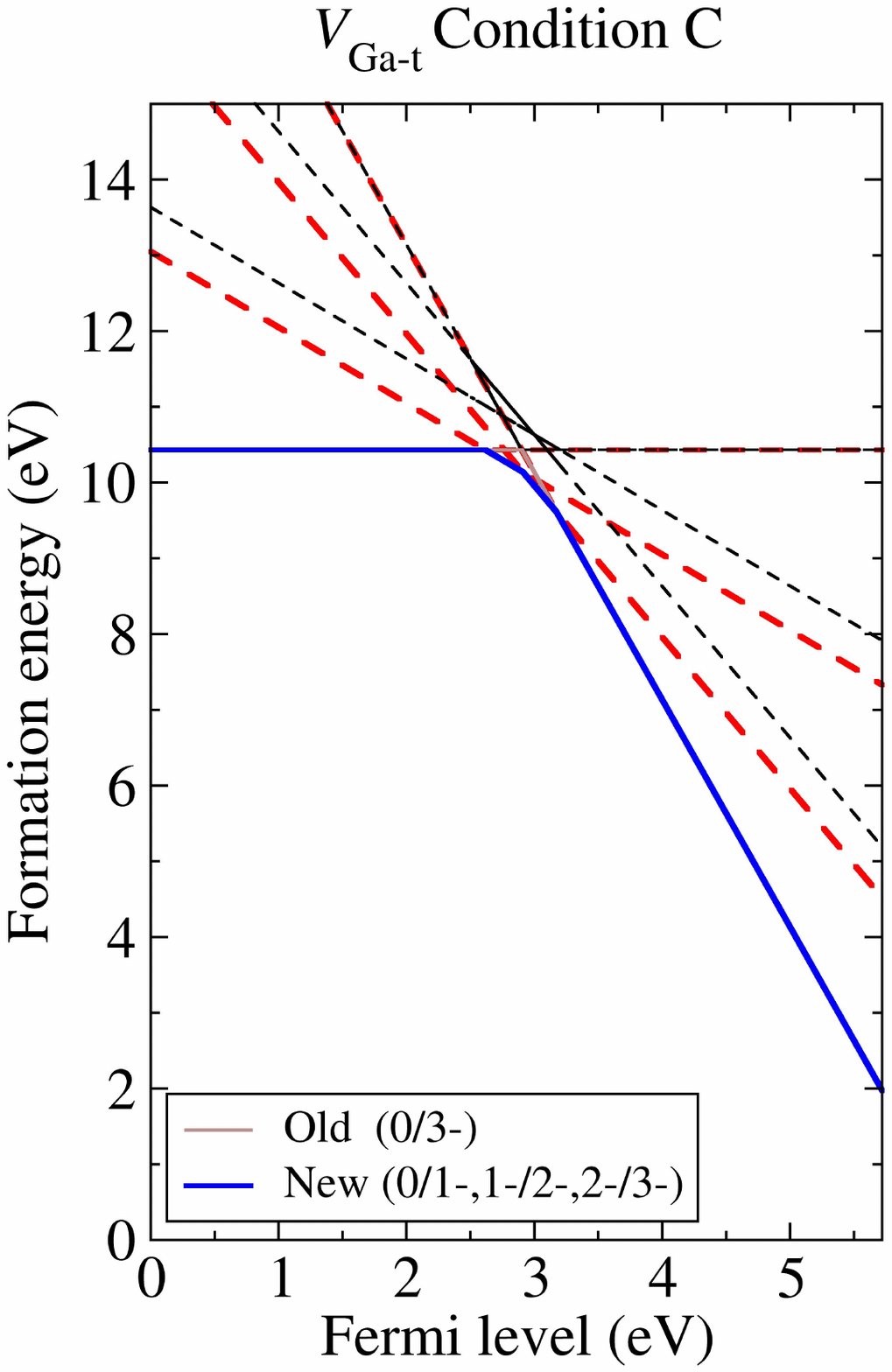}
  \caption{Energies of formation of $V_{\rm Ga-t}$ for different charge states, comparison of fully relaxed structure with previous results from \cite{Dabsamut24}}\label{figefornewgat}
\end{figure}  
\begin{figure*}
(a)  \includegraphics[width=6cm]{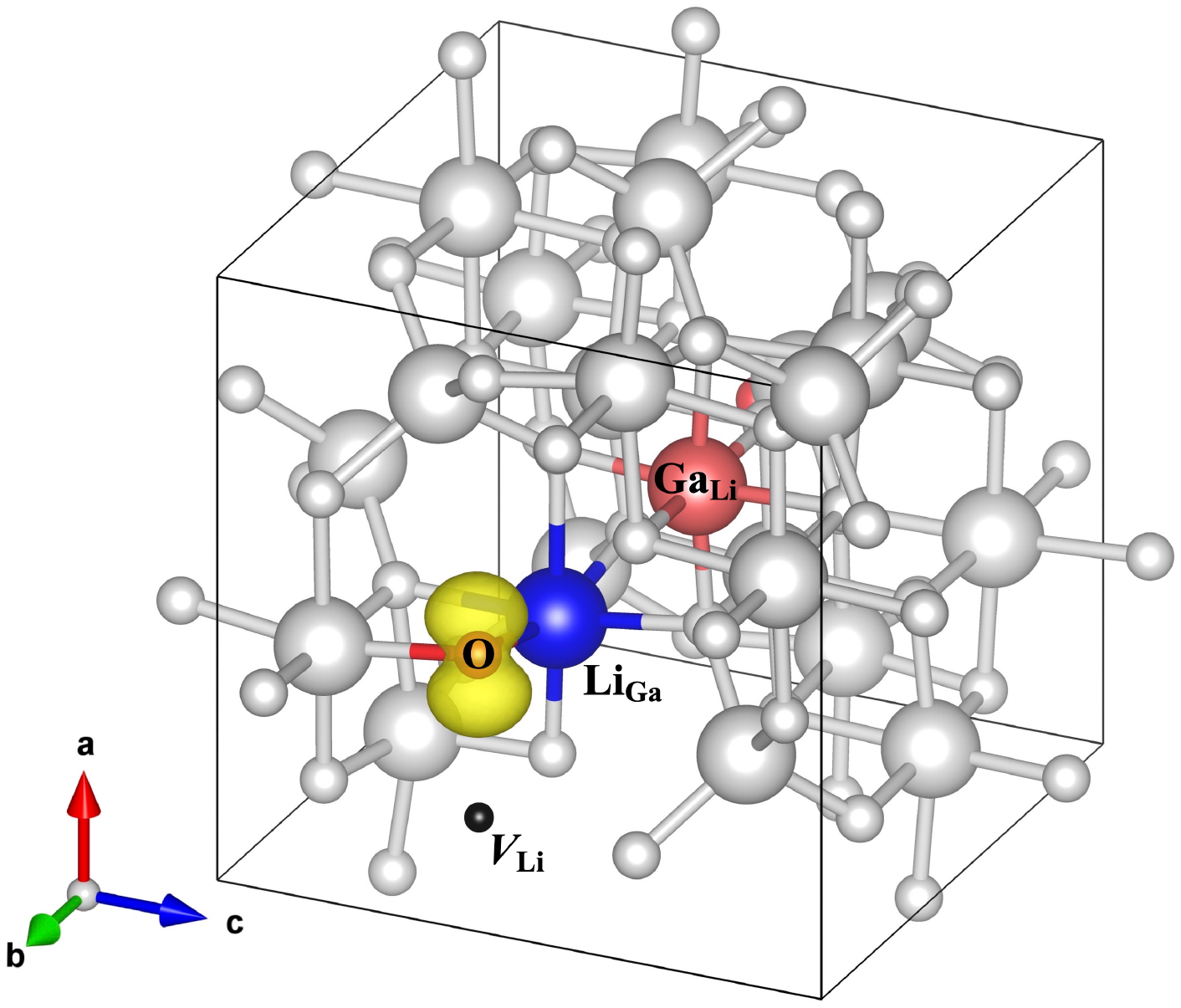}
 (b) \includegraphics[width=8cm]{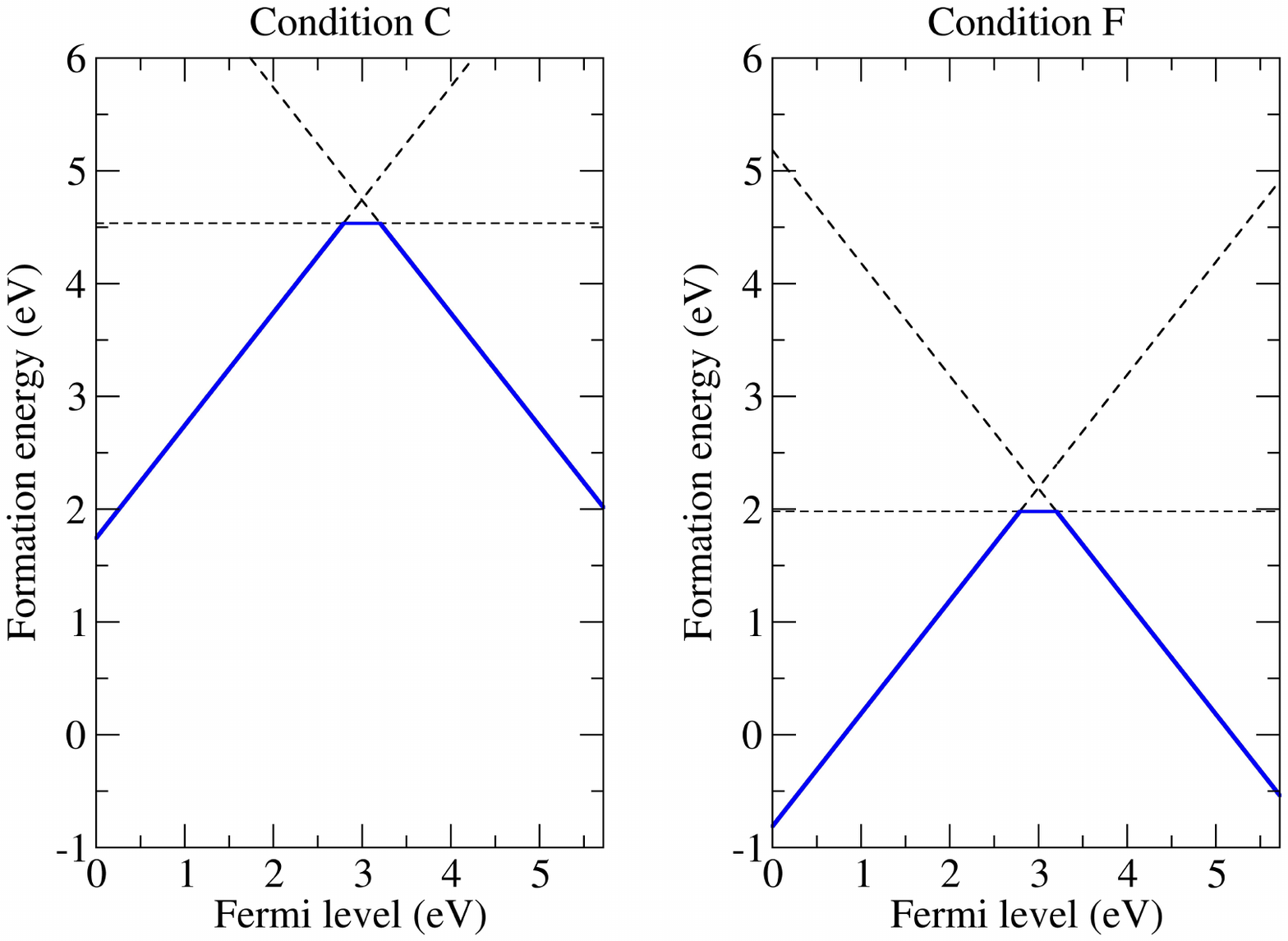}
  \caption{Structure and energy of formation of   $V_{\rm Li}-{\rm Li}_{\rm Ga}-{\rm Ga}_{\rm Li}$ complex.
    Chemical potential condition C is O-poor, F is O-rich.} \label{fig:complex}
\end{figure*}

To gain further insights into the nature of the $V_{\rm Li}$,  we consider the band structure and density of states. The unrelaxed and symmetrically relaxed (non-polaronic) band structure are very similar. In Fig.\ref{figbndvlisym} we show the spin-polarized band structure of the latter. We can see that several minority spin bands are pushed out of the VBM but stay connected to the VBM. The Fermi level in the neutral charge state crosses through the highest of these bands. The gap from the Fermi level to the CBM is about 5.1 eV and is lower than the gap of the perfect crystal in our HSE approximation. This indicates that we should consider the actual gap to be between the top of the majority spin bands (in blue in the figure) and the red states as a defect band. This one-electron picture is consistent with a defect transition level at less than 1 eV above the VBM. On the other hand, in the symmetry broken structure, a single level moves into the gap and remains empty in the neutral charge state. This is consistent with a polaronic state in the gap. In the one-electron picture it lies more than 2 eV above the VBM. 
A similar conclusion follows from the Partial Densities of States (PDOS) shown in Fig.\ref{figpdosvlisym}.

The fact that for the symmetric non-polaronic state, the defect bands pushed in the gap are still connected to the main set of bands may indicate that the unit cell is too small and does not adequately reproduce isolated defect levels. It may thus be important to use a larger cell to adequately gauge the shallow or deep nature of the $V_{\rm Li}$ in the non-polaronic state. Unfortunately, with our current computing resources, it has not yet been possible to perform a HSE calculation for a $2\times2\times2$ supercell of 448 atoms. On the other hand, for the symmetry broken state, the defect level is already clearly pushed in the gap. 

Next, we discuss our revised calculations of the $V_{\rm Ga-t}$. In our previous results of \cite{Dabsamut24} the tetrahedron near a $V_{\rm Gat}$ was already slightly symmetry broken but retained $C_{3v}$ symmetry with one bond length different from the other three. For $q=-1$, three oxygens  were at 2.003 \AA, and one at 2.076 \AA. In the $q=-2$ charge state the atoms moved outward and three were at 2.011 \AA, and one at 2.077 \AA. When allowing lower symmetry we find bond lengths as given in Table\ref{Tab-bonds-VGat}.
The energies of formation under chemical potential condition C (O-poor) is shown in Fig. \ref{figefornewgat} and now
agrees well with Lyons' result of three separate transition levels. Similar results hold for the octahedral Ga vacancy.

\subsection{Native defect complex}\label{sec:complex}
To extend our previous study of the donor acceptor pairs we considered a $V_{\rm Li}-{\rm Li}_{\rm Ga}-{\rm Ga}_{\rm Li}$
complex of neighboring point defects. Since this contains a double donor, a double acceptor and a single acceptor, it is expected to have overall acceptor nature. However, we find it to amphoteric with donor character for Fermi levels near the VBM, a narrow region where it is neutral and then an acceptor type  negative charge state. In other words, it has transition levels $+/0$ at 2.79 eV and a $0/-$ transition at 3.20 eV above the VBM  and fairly high energy of formation under o-poor, cation rich conditions. Furthermore, in the neutral state it has a hole localized on a single O, near the $V_{\rm Li}$.

\subsection{Carbon impurity}\label{sec:carbon}
\begin{figure}
  \includegraphics[width=8cm]{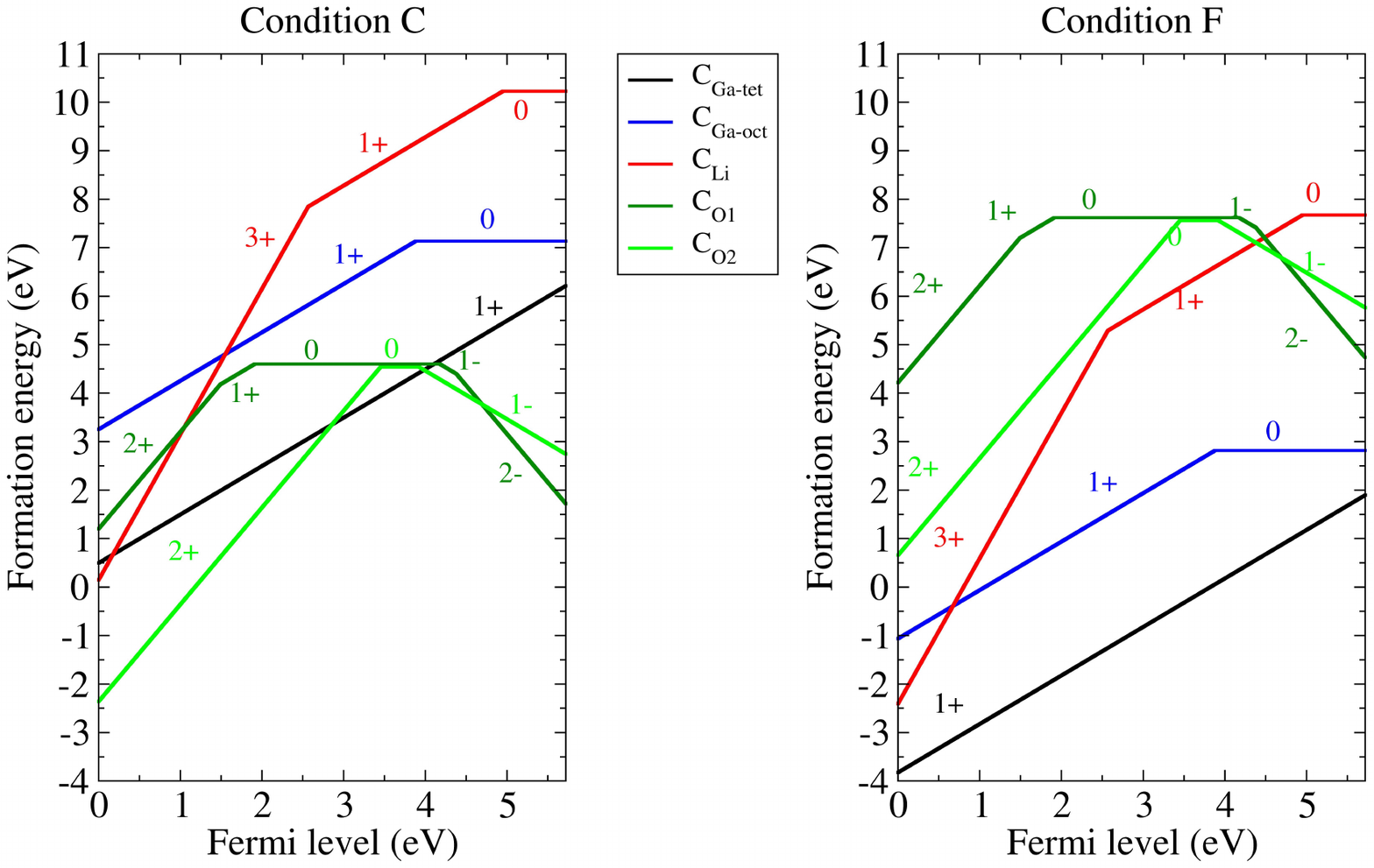}
  \caption{Energies of formation of C substitutional defects for two
    choices of the chemical potential corresponding to C and F in \cite{Dabsamut24} which are respectively O-poor and O-rich.} \label{figcarbon}
\end{figure}

Carbon is a plausible unintentional impurity in chemical vapor deposition (CVD) grown LiGa$_5$O$_8$ because of the organic components in the precursors.
The energies of formation as a function of Fermi level position are shown in Fig.\ref{figcarbon}. We note that the lowest energy of formation occurs for C$_{O2}$ for the positive charge state but this is under the chemical potential condition C of \cite{Dabsamut24}
        which corresponds to the extreme oxygen-poor limit. If we consider a more O-rich condition like F then the energy for formation would be higher by about 3 eV for C$_{\rm O}$ and lower for C$_{\rm Ga}$ by about 4.4 eV and for C$_{\rm Li}$ by 2.5 eV.

The most striking result on the transition levels is that $\mathrm{C}_{\rm Ga\text{-}t}$ behaves as a shallow donor in the sense that its $(1+/0)$ level lies above the quasiparticle gap: $E_t=6.31$~eV and $E_g-E_t=-0.59$~eV. In contrast, the other carbon substitutions introduce deep levels inside the gap. $\mathrm{C}_{\rm Ga\text{-}o}$ has $(1+/0)$ at 3.88 eV, $\mathrm{C}_{\rm Li}$ has a high $(1+/0)$ level at 4.95 eV and a lower $(3+/1+)$ level at 2.57~eV, while $\mathrm{C}_{\rm O1}$ and $\mathrm{C}_{\rm O2}$ exhibit multiple in-gap transitions reflecting their amphoteric character.

\subsection{Configuration diagram results for native defects}\label{sec:config}
\begin{figure}[h]
	 \includegraphics[width=7cm]{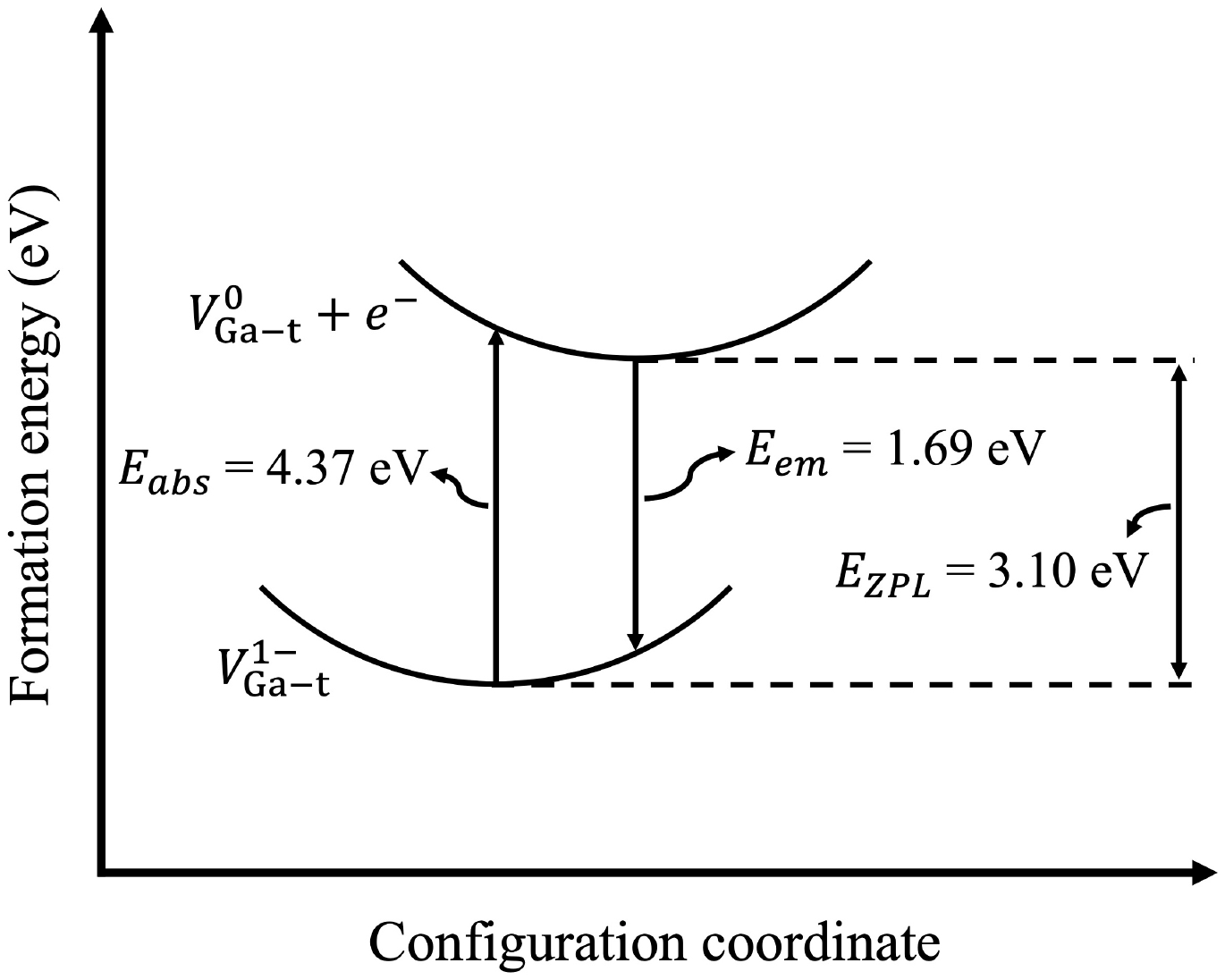}
  \caption{Schematic configuration diagram, identifying $E_{abs}$, $E_{em}$ and zero-phonon line $E_{ZPL}$ for the
    $V_{\rm Ga-t}^0+e^-\rightarrow V_{\rm Ga-t}^{-1}$ case.\label{figscheme}}
   
\end{figure}

As is well known, the transition levels are insufficient to compare with optical data of defects because of the Stokes shift between absorption and emission. To study optical transitions between the band edges and the defect one-electron levels, a common approach is to calculate a defect  configuration diagram of the initial and final charge states. We distinguish two types of transitions. In the first one, the defect is excited by adding an electron to the defect level from the valence band maximum (VBM). 
In this case, we have $E_D^q<E_D^{q-1}$  where $E_D^q$ is the defect formation energy in charge state $q$.
The transition from  charge state $q$ to $q-1$ corresponds to absorption of an electron from the VBM to the defect level and is indicated by $D^q\rightarrow D^{q-1}+h^+$, while the emission is the reverse process $D^{q-1}+h^+\rightarrow D^q$. The second type involves exciting an electron to  the conduction band minimum (CBM) in the absorption step or recombining an electron from the conduction band with a hole in the defect level in the emission step. In this case we have 
$E_D^q<E_D^{q+1}+E_g$ with $E_g$  the band gap and absorption corresponds to the
reaction $D^q\rightarrow D^{q+1}+e^-$ and emission to $D^{q+1}+e^-\rightarrow D^q$. As an example in Fig.\ref{figscheme} we show the configuration diagram of the $V_{\rm Ga-t}$ in the neutral charge state capturing an electron from the CBM.

The assumption behind the use of configuration diagrams is that the time scale of the optical absorption process is too fast to allow the final state to relax and therefore to consider a vertical transition keeping the structure frozen at that of the initial state. We then end up in a vibrationally excited state of the higher energy charge state. Likewise in the emission process, the structure is frozen in the higher energy state and we end up in a vibrationally excited state of the electronic ground state. The line shape of the emission results from the superposition of transitions involving different phonon excited states and
depends on  the overlap of the ground and excited configuration vibrational wave functions and leads to the Huang-Rhys line shape which peaks at the vertical transition energy $E_{PL}$ while the transition energy without any phonon involvement between the two equilibrium energies is called the zero-phonon line $E_{ZPL}$. A full calculation of the optical emission spectra is here not attempted but can be done using the approach developed by Alkauskas \etal\cite{Alkauskas2014}.
A few such absorption and emission energies were reported already by Lyons\cite{Lyons24} but here we present a systematic study of the native defects optical absorption and emission energies complementing our previous study of the equilibrium transition energies, which would correspond to the zero-phonon lines.

\newcommand{\rowsep}{\rule{0pt}{2.4ex}} 
\begin{table*}
	\caption{Thermodynamic transition levels and vertical optical transition energies  for native defects in LiGa$_5$O$_8$.}
	\label{tabresults}
	\begin{ruledtabular}
		\renewcommand{\arraystretch}{1.05} 
		\begin{tabular}{lccc@{\hspace{1.6em}}lccc}
			& \multicolumn{3}{c}{\textbf{Thermodynamic transition levels}} & 
			\multicolumn{4}{c}{\textbf{Vertical optical transition energies}} \\
			\hline
			defect & $q/q'$ & $E_t$ & $E_g-E_t$ &
			transition & $E_{abs}$ & $E_{em}$ & $E_{ZPL}$ \\
			\hline
			
			$V_{\rm Ga-t}$ (symmetric) & $0/3-$ & 2.90 & 2.82 &
			$V_{\rm Ga-t}^{3-}\leftrightarrow V_{\rm Ga-t}^{2-}+e^-$ & 3.78 & 2.55 & 3.21 \\
			&&&& $V_{\rm Ga-t}^{2-}\leftrightarrow V_{\rm Ga-t}^{1-}+e^-$ & 3.37 & 1.97 & 2.72 \\
			&&&& $V_{\rm Ga-t}^{1-}\leftrightarrow V_{\rm Ga-t}^{0}+e^-$  & 3.27 & 1.69 & 2.52 \\
			&&&& $V_{\rm Ga-t}^{3-}+h^+\leftrightarrow V_{\rm Ga-t}^{2-}$ & 3.17 & 1.94 & 2.51 \\
			&&&& $V_{\rm Ga-t}^{2-}+h^+\leftrightarrow V_{\rm Ga-t}^{1-}$ & 3.75 & 2.35 & 3.00 \\
			&&&& $V_{\rm Ga-t}^{1-}+h^+\leftrightarrow V_{\rm Ga-t}^{0}$  & 4.03 & 2.45 & 3.20 \\
			$V_{\rm Ga-t}$ (asymmetric) & $0/1-$ & 2.62 & 3.10 &
			$V_{\rm Ga-t}^{3-}\leftrightarrow V_{\rm Ga-t}^{2-}+e^-$ & 3.78 & 1.14 & 2.54 \\
			& $1-/2-$ & 2.91 & 2.81 & $V_{\rm Ga-t}^{2-}\leftrightarrow V_{\rm Ga-t}^{1-}+e^-$ & 4.05 & 1.46 & 2.81 \\
			& $2-/3-$ & 3.18 & 2.54 & $V_{\rm Ga-t}^{1-}\leftrightarrow V_{\rm Ga-t}^{0}+e^-$  & 4.37 & 1.69 & 3.10 \\
			&&&& $V_{\rm Ga-t}^{3-}+h^+\leftrightarrow V_{\rm Ga-t}^{2-}$ & 4.58 & 1.94 & 3.18 \\
			&&&& $V_{\rm Ga-t}^{2-}+h^+\leftrightarrow V_{\rm Ga-t}^{1-}$ & 4.26 & 1.67 & 2.91 \\
			&&&& $V_{\rm Ga-t}^{1-}+h^+\leftrightarrow V_{\rm Ga-t}^{0}$  & 4.03 & 1.35 & 2.62 \\
			\hline
			
			 $V_{\rm Ga-o}$ (symmetric)& $0/1-$  & 1.84 & 3.88 &
			$V_{\rm Ga-o}^{3-}\leftrightarrow V_{\rm Ga-o}^{2-}+e^-$ & 3.82 & 2.15 & 3.01 \\
			 & $1-/3-$ & 2.92 & 2.80 &
			$V_{\rm Ga-o}^{2-}\leftrightarrow V_{\rm Ga-o}^{1-}+e^-$ & 3.42 & 1.57 & 2.59 \\
			&&&& $V_{\rm Ga-o}^{1-}\leftrightarrow V_{\rm Ga-o}^{0}+e^-$  & 4.94 & 2.69 & 3.88 \\
			&&&& $V_{\rm Ga-o}^{3-}+h^+\leftrightarrow V_{\rm Ga-o}^{2-}$ & 3.57 & 1.90 & 2.71 \\
			&&&& $V_{\rm Ga-o}^{2-}+h^+\leftrightarrow V_{\rm Ga-o}^{1-}$ & 4.15 & 2.30 & 3.13 \\
			&&&& $V_{\rm Ga-o}^{1-}+h^+\leftrightarrow V_{\rm Ga-o}^{0}$  & 3.03 & 0.78 & 1.84 \\
						 $V_{\rm Ga-o}$ (asymmetric)& $0/1-$  & 2.20 & 3.52 &
			$V_{\rm Ga-o}^{3-}\leftrightarrow V_{\rm Ga-o}^{2-}+e^-$ & 3.82 & 1.16 & 2.65 \\
			& $1-/2-$ & 2.77 & 2.95 &
			$V_{\rm Ga-o}^{2-}\leftrightarrow V_{\rm Ga-o}^{1-}+e^-$ & 4.16 & 1.57 & 2.95 \\
			& $2-/3-$  & 3.07 & 2.65 & $V_{\rm Ga-o}^{1-}\leftrightarrow V_{\rm Ga-o}^{0}+e^-$  & 4.94 & 2.24 & 3.52 \\
			&&&& $V_{\rm Ga-o}^{3-}+h^+\leftrightarrow V_{\rm Ga-o}^{2-}$ & 4.56 & 1.90 & 3.07 \\
			&&&& $V_{\rm Ga-o}^{2-}+h^+\leftrightarrow V_{\rm Ga-o}^{1-}$ & 4.15 & 1.56 & 2.77 \\
			&&&& $V_{\rm Ga-o}^{1-}+h^+\leftrightarrow V_{\rm Ga-o}^{0}$  & 3.48 & 0.78 & 2.20 \\
			\hline
			
			$V_{\rm Li}$ (symmetric) & $0/1-$ & 0.74 & 4.98 &
			$V_{\rm Li}^{1-}\leftrightarrow V_{\rm Li}^{0}+e^-$       & 5.40 & 4.52 & 4.98 \\
			&&&& $V_{\rm Li}^{0}+h^+\leftrightarrow V_{\rm Li}^{1-}$  & 1.20 & 0.32 & 0.74 \\
			$V_{\rm Li}$ (asymmetric) & $0/1-$ &1.58 & 4.14 &
			$V_{\rm Li}^{1-}\leftrightarrow V_{\rm Li}^{0}+e^-$       & 5.40 & 2.69 & 4.14 \\
			&&&& $V_{\rm Li}^{0}+h^+\leftrightarrow V_{\rm Li}^{1-}$  & 3.03 & 0.32 & 1.58 \\
			\hline
			
			$V_{\rm O1}$ & $2+/0$ & 3.80 & 1.92 &
			$V_{\rm O1}^{1+}\leftrightarrow V_{\rm O1}^{2+}+e^-$      & 3.53 & 0.40 & 1.89 \\
			&&&& $V_{\rm O1}^{0}\leftrightarrow V_{\rm O1}^{1+}+e^-$   & 3.23 & 0.66 & 1.94 \\
			&&&& $V_{\rm O1}^{1+}+h^+\leftrightarrow V_{\rm O1}^{2+}$  & 5.34 & 3.19 & 3.83 \\
			&&&& $V_{\rm O1}^{0}+h^+\leftrightarrow V_{\rm O1}^{1+}$   & 5.06 & 2.49 & 3.78 \\
			\hline
			
			$V_{\rm O2}$ & $2+/0$ & 3.87 & 1.85 &
			$V_{\rm O2}^{1+}\leftrightarrow V_{\rm O2}^{2+}+e^-$      & 3.65 & $-$0.11 & 1.65 \\
			&&&& $V_{\rm O2}^{0}\leftrightarrow V_{\rm O2}^{1+}+e^-$   & 3.59 & 0.66 & 2.06 \\
			&&&& $V_{\rm O2}^{1+}+h^+\leftrightarrow V_{\rm O2}^{2+}$  & 5.83 & 2.07 & 4.07 \\
			&&&& $V_{\rm O2}^{0}+h^+\leftrightarrow V_{\rm O2}^{1+}$   & 5.06 & 2.13 & 3.66 \\
			\hline
			
			Li$_{\rm Ga-t}$ (symmetric)& $0/2-$ & 1.74 & 3.98 &
			Li$_{\rm Ga-t}^{2-}\leftrightarrow$Li$_{\rm Ga-t}^{1-}+e^-$ & 4.85 & 3.45 & 4.26 \\
			&&&& Li$_{\rm Ga-t}^{1-}\leftrightarrow$Li$_{\rm Ga-t}^{0}+e^-$ & 4.41 & 2.91 & 3.70 \\
			&&&& Li$_{\rm Ga-t}^{2-}+h^+\leftrightarrow$Li$_{\rm Ga-t}^{1-}$ & 2.27 & 0.87 & 1.46 \\
			&&&& Li$_{\rm Ga-t}^{1-}+h^+\leftrightarrow$Li$_{\rm Ga-t}^{0}$  & 2.81 & 1.31 & 2.02 \\
			Li$_{\rm Ga-t}$ (asymmetric) & $0/1-$ & 2.06 & 3.66 &
			Li$_{\rm Ga-t}^{2-}\leftrightarrow$Li$_{\rm Ga-t}^{1-}+e^-$ & 4.85 & 2.62 & 4.26 \\
			& $1-/2-$ & 2.30 & 3.42 & Li$_{\rm Ga-t}^{1-}\leftrightarrow$Li$_{\rm Ga-t}^{0}+e^-$ & 4.20 & 2.99 & 3.70 \\
			&&&& Li$_{\rm Ga-t}^{2-}+h^+\leftrightarrow$Li$_{\rm Ga-t}^{1-}$ & 3.10 & 0.87 & 1.46 \\
			&&&& Li$_{\rm Ga-t}^{1-}+h^+\leftrightarrow$Li$_{\rm Ga-t}^{0}$  & 2.73 & 1.52 & 2.02 \\
			\hline
			
			Li$_{\rm Ga-o}$ (symmetric) & $0/2-$ & 2.10 & 3.62 &
			Li$_{\rm Ga-o}^{2-}\leftrightarrow$Li$_{\rm Ga-o}^{1-}+e^-$ & 4.75 & 2.78 & 3.89 \\
			&&&& Li$_{\rm Ga-o}^{1-}\leftrightarrow$Li$_{\rm Ga-o}^{0}+e^-$ & 4.31 & 2.34 & 3.36 \\
			&&&& Li$_{\rm Ga-o}^{2-}+h^+\leftrightarrow$Li$_{\rm Ga-o}^{1-}$ & 2.94 & 0.97 & 1.83 \\
			&&&& Li$_{\rm Ga-o}^{1-}+h^+\leftrightarrow$Li$_{\rm Ga-o}^{0}$  & 3.38 & 1.41 & 2.36 \\
			Li$_{\rm Ga-o}$ (asymmetric) & $0/1-$ & 1.93 & 3.79 &
			Li$_{\rm Ga-o}^{2-}\leftrightarrow$Li$_{\rm Ga-o}^{1-}+e^-$ & 4.75 & 2.34 & 3.89 \\
			& $1-/2-$ & 2.27 & 3.45 & Li$_{\rm Ga-o}^{1-}\leftrightarrow$Li$_{\rm Ga-o}^{0}+e^-$ & 4.64 & 2.34 & 3.36 \\
			&&&& Li$_{\rm Ga-o}^{2-}+h^+\leftrightarrow$Li$_{\rm Ga-o}^{1-}$ & 3.38 & 0.97 & 1.83 \\
			&&&& Li$_{\rm Ga-o}^{1-}+h^+\leftrightarrow$Li$_{\rm Ga-o}^{0}$  & 3.38 & 1.08 & 2.36 \\
			
		\end{tabular}
	\end{ruledtabular}
\end{table*}

Table \ref{tabresults} lists the transition levels, the  absorption and emission and zero-phonon line
levels involving the VBM and CBM for native defects.
We consider here all possible vertical transitions but not all are equally relevant. 
This depends on the equilibrium Fermi level and  hence the charge state of the defect. 
For example, the  $V_{\rm Ga-t}$ will be in the $3-$ charge state  if the Fermi level is above the $3-/2-$ transition level,  so it cannot  acquire any more electrons but in an optical absorption process it can release its electron to the CBM by transitioning to the $2-$ charge state. On the other hand the corresponding emission process involving the CBM is unlikely because  for low excitation level, only few $V_{\rm Ga-t} $ will be in the $2-$ charge state, only those that have already being excited. 
However, it can cause luminescence by recombination of  the electron in the defect with a hole in the VBM.
With our  revised transition levels for $V_{\rm Ga-t}$ and $V_{\rm Ga-o}$, we find fair agreement with the results of Lyons for $V_{\rm Ga}$ although he did not distinguish between the two types of Ga-vacancies.   He finds $E_{abs}=2.93$ eV, $E_{PL}=1.27$ eV and the $E_{ZPL}=2.19$ eV for the transitions between the CBM and the $2-/3-$ transition level, whereas we find  $E_{abs}=3.78$ eV, $E_{PL}=1.14$ eV and the $E_{ZPL}=2.54$ eV for the $V_{\rm Ga-t}$ and  $E_{abs}=3.82$ eV, $E_{PL}=1.16$ eV and the $E_{ZPL}=2.65$ eV. The remaining
discrepancy results from our larger band gap of 5.7 instead of 5.3 eV. 
For the $V_{\rm Li}$, Lyons gives $E_{PL}=3.08$ eV, $E_{abs}=4.84$ eV and $E_{ZPL}=4.08$ eV for transitions between the conduction band and the $0/-$ level, we  have now a deeper level in the symmetry-broken state and thus there remain
some additional differences. 

Table~\ref{tabCsubs} summarizes both the thermodynamic charge-transition levels and the corresponding vertical optical transitions for substitutional carbon on the tetrahedral and octahedral Ga sites, on Li, and on the two inequivalent O sites. The optical transitions related to C will be discussed in the next section.

        \begin{table*}
	\caption{Thermodynamic transition levels and vertical optical transition energies for carbon substitution defects in LiGa$_5$O$_8$.}
	\label{tabCsubs}
	\begin{ruledtabular}
		\renewcommand{\arraystretch}{1.05}
		\begin{tabular}{lccc@{\hspace{1.6em}}lccc}
			& \multicolumn{3}{c}{\textbf{Thermodynamic transition levels}} &
			\multicolumn{4}{c}{\textbf{Vertical optical transition energies}} \\
			\hline
			defect & $q/q'$ & $E_t$ & $E_g-E_t$ &
			transition & $E_{abs}$ & $E_{em}$ & $E_{ZPL}$ \\
			\hline
			
			C$_{\rm Ga-t}$ & $(1+/0)$ & 6.31 & $-$0.59 &
			C$_{\rm Ga-t}^{0}\leftrightarrow \mathrm{C}_{\rm Ga-t}^{1+}+e^-$ & $-$0.33 & $-$0.85 & $-$0.59 \\
			&&&& $\mathrm{C}_{\rm Ga-t}^{0}+h^+\leftrightarrow \mathrm{C}_{\rm Ga-t}^{1+}$ & 6.57 & 6.05 & 6.31 \\
			\hline
			
			C$_{\rm Ga-o}$ & $(1+/0)$ & 3.88 & 1.84 &
			C$_{\rm Ga-o}^{0}\leftrightarrow \mathrm{C}_{\rm Ga-o}^{1+}+e^-$ & 3.82 & $-$0.48 & 1.84 \\
			&&&& $\mathrm{C}_{\rm Ga-o}^{0}+h^+\leftrightarrow \mathrm{C}_{\rm Ga-o}^{1+}$ & 6.20 & 1.90 & 3.88 \\
			\hline
			
			$\mathrm{C}_{\rm Li}$ & $(3+/1+)$ & 2.57 & 3.15 &
			$\mathrm{C}_{\rm Li}^{2+}\leftrightarrow \mathrm{C}_{\rm Li}^{3+}+e^-$ & 4.43 & 0.99 & 2.77 \\
			& $(1+/0)$ & 4.95 & 0.77 &
			$\mathrm{C}_{\rm Li}^{1+}\leftrightarrow \mathrm{C}_{\rm Li}^{2+}+e^-$ & 4.69 & 1.96 & 3.54 \\
			&&&& $\mathrm{C}_{\rm Li}^{0}\leftrightarrow \mathrm{C}_{\rm Li}^{1+}+e^-$ & 2.14 & $-$0.64 & 0.77 \\
			&&&& $\mathrm{C}_{\rm Li}^{2+}+h^+\leftrightarrow \mathrm{C}_{\rm Li}^{3+}$ & 4.73 & 1.29 & 2.95 \\
			&&&& $\mathrm{C}_{\rm Li}^{1+}+h^+\leftrightarrow \mathrm{C}_{\rm Li}^{2+}$ & 3.76 & 1.03 & 2.18 \\
			&&&& $\mathrm{C}_{\rm Li}^{0}+h^+\leftrightarrow \mathrm{C}_{\rm Li}^{1+}$ & 6.36 & 3.58 & 4.95 \\
			\hline
			
			$\mathrm{C}_{\rm O1}$ & $(2+/1+)$ & 1.49 & 4.23 &
			$\mathrm{C}_{\rm O1}^{2-}\leftrightarrow \mathrm{C}_{\rm O1}^{1-}+e^-$ & 2.39 & 0.43 & 1.34 \\
			& $(1+/0)$ & 1.91 & 3.81 &
			$\mathrm{C}_{\rm O1}^{1-}\leftrightarrow \mathrm{C}_{\rm O1}^{0}+e^-$ & 1.12 & 0.42 & 1.54 \\
			& $(0/1-)$ & 4.18 & 1.54 &
			$\mathrm{C}_{\rm O1}^{1+}\leftrightarrow \mathrm{C}_{\rm O1}^{2+}+e^-$ & 5.23 & 2.95 & 4.23 \\
			& $(1-/2-)$ & 4.38 & 1.34 &
			$\mathrm{C}_{\rm O1}^{0}\leftrightarrow \mathrm{C}_{\rm O1}^{1+}+e^-$ & 4.74 & 2.71 & 3.81 \\
			&&&& $\mathrm{C}_{\rm O1}^{2-}+h^+\leftrightarrow \mathrm{C}_{\rm O1}^{1-}$ & 5.29 & 3.33 & 4.38 \\
			&&&& $\mathrm{C}_{\rm O1}^{1-}+h^+\leftrightarrow \mathrm{C}_{\rm O1}^{0}$ & 5.30 & -1.12 & 4.18 \\
			&&&& $\mathrm{C}_{\rm O1}^{1+}+h^+\leftrightarrow \mathrm{C}_{\rm O1}^{2+}$ & 2.77 & 0.49 & 1.49 \\
			&&&& $\mathrm{C}_{\rm O1}^{0}+h^+\leftrightarrow \mathrm{C}_{\rm O1}^{1+}$ & 3.01 & 0.98 & 1.91 \\
			\hline
			
			$\mathrm{C}_{\rm O2}$ & $(2+/0)$ & 3.45 & 2.27 &
			$\mathrm{C}_{\rm O2}^{2-}\leftrightarrow \mathrm{C}_{\rm O2}^{1-}+e^-$ & 0.82 & -1.12 & -0.17 \\
			& $(0/1-)$ & 3.92 & 1.80 &
			$\mathrm{C}_{\rm O2}^{1-}\leftrightarrow \mathrm{C}_{\rm O2}^{0}+e^-$ & 1.28 & 0.52 & 1.80 \\
			&&&& $\mathrm{C}_{\rm O2}^{1+}\leftrightarrow \mathrm{C}_{\rm O2}^{2+}+e^-$ & 5.67 & -0.68 & 1.09 \\
			&&&& $\mathrm{C}_{\rm O2}^{0}\leftrightarrow \mathrm{C}_{\rm O2}^{1+}+e^-$ & 4.43 & 2.38 & 3.45 \\
			&&&& $\mathrm{C}_{\rm O2}^{2-}+h^+\leftrightarrow \mathrm{C}_{\rm O2}^{1-}$ & 6.84 & 4.90 & 5.89 \\
			&&&& $\mathrm{C}_{\rm O2}^{1-}+h^+\leftrightarrow \mathrm{C}_{\rm O2}^{0}$ & 5.20 & -1.28 & 3.92 \\
			&&&& $\mathrm{C}_{\rm O2}^{1+}+h^+\leftrightarrow \mathrm{C}_{\rm O2}^{2+}$ & 6.40 & 0.05 & 4.63 \\
			&&&& $\mathrm{C}_{\rm O2}^{0}+h^+\leftrightarrow \mathrm{C}_{\rm O2}^{1+}$ & 3.34 & 1.29 & 2.27 \\

		\end{tabular}
	\end{ruledtabular}
\end{table*}

\subsection{Discussion of experimental data}\label{sec:expt}
In \cite{Dabsamut24} we found that the equilibrium Fermi level dictated by the charge neutrality condition is at $\sim$3.25 eV above the VBM when considering only native defects and slightly higher at 3.54 if we also consider various donor acceptor pair complexes. Note that even with the now lower $V_{\rm Li}^0$  and hence higher $-1/0$ transition level, this does not change the $V_{\rm Li}^{1-}$ state and hence its intersection with the Ga$_{\rm Li}^{2+}$ which determines the Fermi level. 
To predict likely optical signatures resulting from the native defects, we assume that the defect concentrations  stay close to equilibrium even under the excitation process of cathodo- or photoluminescence. 
This is  indeed expected if the  optical excitation level has low intensity. The excitation process however, is assumed to provide electrons in the CBM and holes at the VBM to recombine with.  
This means that the acceptor type defects are all in their most negative charge state and the $V_{\rm O}$ are in their most positive $2+$ charge state. Thus, for radiative emission  processes, the acceptors can only have transitions to the VBM while the oxygen vacancy donors can only capture electrons from the CBM but not emit holes to the VBM. This means emission peaks are expected at 1.94 eV, 1.90 eV for $V_{\rm Ga-t}$ and $V_{\rm Ga-o}$, 0.32 eV for $V_{\rm Li}$ and 0.87 eV, 0.97 eV for Li$_{\rm Ga-t}$ and Li$_{\rm Ga-o}$.
Considering the symmetry broken $V_{\rm Gat}$, the $3-/2-$ now lies at 3.18 eV, closer to the equilibrium Fermi level, there is a small probability it could catch an electron from the CBM in its $2-$ state with an emission at 1.14 eV. Similar for the $V_{\rm Ga-o}$, with  $3-/2-$ at 3.07 giving rise to emission when capturing an electron from the CBM at 1.16 eV. CL lines at these energies have not been observed. 
On the other hand, $V_{\rm Ga}$ defects have high energy of formation and are expected to have only very low concentrations, especially in somewhat Ga-rich conditions. 

Experimentally, the samples in which p-type conduction were found to be oxygen deficient and somewhat Ga rich. Nonetheless, we found that at a growth temperature of 1200K, under conditions E, F in \cite{Dabsamut24}, we are still closer to the O-rich than the extreme O-poor limit where we would be close to precipitating pure Ga droplets, which have not been observed. If we consider rather poor O-conditions closer to  C or D, the equilibrium Fermi level would move higher in the gap but would still be pinned by the intersection of $V_{\rm Li}$ and Ga$_{\rm Li}$. Figure  S4 in the supplemental information of \cite{Dabsamut24}  indicates that the equilibrium Fermi level could then lie near 4 eV and thus just above the $V_{\rm O}$ transition levels.
In that case, the $V_{\rm O}$ could have transitions to the VBM which would occur at 2.49 eV and 2.13 eV for O1 and O2 respectively. 
$V_{\rm Ga}$ in their $q=-3$ charge state at this Fermi level position  would still have energies of formation of 5 eV or higher and are unlikely to be present in the samples.

In cathodoluminescence experiments\cite{Kaitian25} on LiGa$_5$O$_8$ grown on a  sapphire substrates and capped by a Ga$_2$O$_3$ layer, peaks are found at $\sim$ 1.8 eV and 3 eV with a shoulder at 3.5-3.6 eV. The peak at 1.8 eV is rather sharp and close to
a known internal transition of  Cr $d$-states which is a likely impurity in  both the CVD grown Ga$_2$O$_3$ and LiGa$_5$O$_8$ from the stainless steel parts of the CVD growth chamber\cite{Kaitian25}. The peak at 3 eV is  significantly broader and stretches from about 2 to 4.5 eV. In the case of a sapphire substrate a distinct peak is also present near 2.5 eV. This feature  is also present as a shoulder in the case of a Ga$_2$O$_3$ substrate and is found to be one of the few peaks that is dominant in the LiGa$_5$O$_8$ region of the sample by depth resolved CL analysis. We therefore tentatively assign this feature to the O-vacancies recombining with the VBM. This would however require that the Fermi level lies high in the  conduction band which is incompatible with p-type conditions.
This assignment could be further tested by examining its behavior under oxidizing or reducing annealing treatments.
While the $V_{\rm Ga}$ predicted luminescence peaks at about 1.9 eV could contribute to the sharp peak at 1.85 eV this is somewhat unlikely given that the $V_{\rm Ga}$ have high energy of formation even in their negative charge state and with the Fermi level this high in the gap.
The 3 eV peak is strong in the Ga$_2$O$_3$ region \cite{Gao2018} and may thus not be intrinsic to LiGa$_5$O$_8$. Our analysis indicates that none of the native defects can be responsible for CL emission in this energy range. The shoulder at 3.5 eV which corresponds to 339 nm was previously also associated with a Cr related transition \cite{Kaitian25}.

	For the C-impurities, the optical interpretation follows the same logic as for the native defects: once a charge state is selected by the Fermi level, one may consider optical transitions associated with electron capture from the conduction band or hole capture from the valence band. Intermediate charge states that are not thermodynamically stable may still be accessed transiently under excitation.
	For $\mathrm{C}_{\rm Ga\text{-}t}$, electron-capture transitions are not meaningful within the present configuration-coordinate construction because the donor level lies above the gap; this is reflected by negative values for the corresponding vertical transition.

        We now discuss if these defects could be responsible for some of the observed CL peaks. If we first assume an equilibrium Fermi level at 3.25 eV, the C$_{\rm Ga-t}$ and C$_{\rm Ga-o}$  would both be in the positive charge state and cannot emit by recombining an electron with the VBM because they don't have any electrons. They can also not capture an electron from the CBM because their neutral charge state
        at the structure of the positive charge state lies  above the energy of the positive charge state  plus a conduction electron.
        The C$_{\rm O}$ would both be in a neutral charge state and being amphoteric can interact with both the VBM and CBM. Capturing an electron from the CBM to make a $q=-1$ state would occur at 0.42, 0.5 eV for O1, O2 respectively. On the other hand recombining one of their electrons with  a hole at the VBM would occur at 0.98 eV, 1.29 eV for O1, O2 respectively. If the Fermi level lies above 4 eV these C$_{\rm O}$
        could possibly occur in a $q=-1$ charge state but their emission processes with the CBM or VBM then either are very small or negative and thus forbidden. Finally, consider the C$_{\rm Li}$. This defect  would be in a $q=+1$ charge state. It cannot capture an electron from the conduction band but can emit by recombining with a hole forming a 2+ charge state, which would emit light at 1.03 eV. Only if the Fermi level
        is as high as 4.95 eV it could be in a neutral charge state and the recombination of this state with a hole from the VBM would occur at 3.58 eV. So this last process seems a remote possibility to explain the 3.5 eV observed CL but it requires the Fermi level to lie quite high in the gap at almost 5 eV. If the Fermi level is indeed that high, then the C$_{\rm O1}$ would be in the $q=-2$ charge state and could then recombine with a hole at the VBM by emitting light at 3.33 eV,  which would also contribute to the broad CL peak centered at 3 eV. These results suggest that C impurities are possibly contributing to the CL broad peak at 3-3.5 eV but only if the Fermi level is high in the gap.

\section{Conclusion}
\label{sec:conclusion}
In this paper we critically re-examined our previous results for the defects in LiGa$_5$O$_8$. First, we performed configuration diagram calculations of the  optical transitions between different charge states by transitions to both the valence and conduction band edges.
Second, we found that upon full symmetry breaking relaxation of the Li-vacancy, we now find a polaronic state with the hole localized on a single oxygen and deeper than we had before, in qualitative agreement with the results of Lyons \cite{Lyons24}. However, this also suggests that the $V_{\rm Li}$ could be a dual nature defect with a shallower symmetric and deeper symmetry broken state. 
The non-polaronic state found thus far is not sufficiently shallow to explain p-type doping  but this might possibly be a limitation of the size of the supercell. However, even  if $V_{\rm Li}$ turns out to have a shallower level in the dilute limit, it does not appear to be protected from the deep polaronic state by a barrier and is not metastable. Furthermore, compensation of any shallow  acceptor by the shallow donor Ga$_{\rm Li}$, which has strongly negative formation energy for Fermi levels  near the VBM would produce insulating behavior in  equilibrium. It thus appears that the observed p-type behavior cannot stem from LiGa$_5$O$_8$  but must be related to a secondary phase in the nanostructure of unknown type.

The optical luminescence transitions were calculated and discussed in the context of the expected equilibrium Fermi level position under stoichiometric and somewhat oxygen deficient conditions. We find that  the only match with presently available cathodoluminescence data is for oxygen vacancies  giving luminescence by recombining with holes in the VBM and these would lead to a peak at $\sim2.5$  eV but this would only occur if the Fermi level lies above the $2+/0$  transition levels of the O vacancies, which is above 4 eV above the VBM, which is  incompatible with p-type doping. While $V_{\rm Ga}$ transitions to the VBM near 1.9 eV could in principle contribute to a sharp peak seen at 1.85 eV, this is deemed unlikely because of the high energy of formation of Ga vacancies.

We also studied carbon impurity formation energies, transition levels and related optical transitions for various substitutional sites.
We find that the site selectivity depends strongly on the stoichiometry or chemical potentials. Under strongly O-poor conditions the O sites are favored while under somewhat richer O conditions, cation sites are favored.  We find that C on tetrahedral Ga sites is a shallow donor while on the other sites, it is a deep donor defect and on oxygen sites amphoteric. We found no  optical transitions that match known luminescence lines if  the equilibrium Fermi level  lies in the middle of the gap. Only with a Fermi level rather high in the gap, which would only occur under rather O-poor conditions, we find a possible contribution for C$_{\rm O1}$ and C$_{\rm Li}$ luminescence by recombination with holes in the VBM  to the observed broad luminescence around 3.3-3.5 eV.

\section{ACKNOWLEDGMENTS}
This work was funded by the US Air Force Office of Scientific Research (AFOSR), Program Manager Dr. Ali Sayir, on under grant number FA9550-26-1-0005. We thank Profs. Hongping Zhao and Leonard J. Brillson 
and Carlos DeLeon for useful discussions about their  cathodoluminescence  studies. 
This work utilized the high-performance computing resources provided by the Core Facility for Advanced Research Computing at Case Western Reserve University.

\section{Data availability}
The data that support the findings of this study are available within the article.

\bibliography{ga2o3,ligao2,liga5o8,dft,defects}
\end{document}